\documentclass[a4paper]{appolb}

\usepackage{hyperref}
\usepackage{amsfonts,amssymb,amsmath,amsthm,cite}
\usepackage{graphicx}

\DeclareMathOperator{\Res}{Res}         

\newtheorem{assumption}{Assumption}[section]
\newtheorem{theorem}[assumption]{Theorem}
\newtheorem{corollary}[assumption]{Corollary}

\newtheorem{lemma}[assumption]{Lemma}
\newtheorem{definition}[assumption]{Definition}
\newtheorem{prop}[assumption]{Proposition}
\newtheorem{remark}[assumption]{Remark}

\newcommand{\A}{\mathcal{A}}              
\renewcommand{\a}{\alpha}                    
\newcommand{\C}{\mathbb{C}}              
\newcommand{\DD}{\mathcal{D}}           
\renewcommand{\H}{\mathcal{H}}          
\newcommand{\K}{\mathcal{K}}             
\renewcommand{\L}{\mathcal{L}}          

\newcommand{\N}{\mathbb{N}}            
\newcommand{\OO}{\mathcal{O}}

\newcommand{\Q}{\mathbb{Q}}             
\newcommand{\R}{\mathbb{R}}             

\newcommand{\set}[1]{\{\,#1\,\}}              
\renewcommand{\SS}{\mathcal{S}}        
\newcommand{\T}{\mathbb{T}}                
\newcommand{\U}{\mathcal{U}}              
\newcommand{\x}{\times}                      
\newcommand{\Z}{\mathbb{Z}}                 

\newcommand{\bb}{\begin{eqnarray}}
\newcommand{\ee}{\end{eqnarray}}
\newcommand{\eee}{\nonumber\end{eqnarray}}

\newbox\ncintdbox \newbox\ncinttbox
\setbox0=\hbox{$-$} \setbox2=\hbox{$\displaystyle\int$}
\setbox\ncintdbox=\hbox{\rlap{\hbox
    to \wd2{\kern-.1em\box2\relax\hfil}}\box0\kern.1em}
\setbox0=\hbox{$\vcenter{\hrule width 4pt}$}
\setbox2=\hbox{$\textstyle\int$}
\setbox\ncinttbox=\hbox{\rlap{\hbox to \wd2{\kern-.14em\box2\relax\hfil}}\box0\kern.1em}
\newcommand{\ncint}{\mathop{\mathchoice{\copy\ncintdbox}
    {\copy\ncinttbox}{\copy\ncinttbox}
    {\copy\ncinttbox}}\nolimits}

\begin{document}
\pagestyle{plain}
\newcount\eLiNe\eLiNe=\inputlineno\advance\eLiNe by -1

\title{$\kappa$-Deformation and Spectral Triples
\author{
B. Iochum \footnote{Talk presented by B. I. at the conference ``Geometry and Physics in Cracow", September 21-25, 
2010, iochum@cpt.univ-mrs.fr}, 
T. Masson, 
T. Sch\"ucker 
\& A. Sitarz 
\thanks{The Smoluchowski Institute of Physics, Jagiellonian University, Reymonta 4, 30-059 Krak\'ow, Poland, Partially supported by MNII grant 189/6.PRUE/2007/7 and N 201 1770 33}
\address{
Centre de Physique Th\'eorique
\thanks{Unit\'e Mixte de Recherche du CNRS et des Universit\'es Aix-Marseille I, Aix-Marseille II et de l'Universit\'e 
du Sud Toulon-Var},\\
CNRS--Luminy, Case 907
13288 Marseille Cedex 9
FRANCE
}}}
\maketitle
\begin{abstract}
The aim of the paper is to answer the following question: does $\kappa$-defor\-mation fit into the framework of 
noncommutative geometry in the sense of spectral triples? \\
Using a compactification of time, we get a discrete version of $\kappa$-Minkowski deformation via 
$C^*$-algebras of groups. 
The dynamical system of the underlying groups (including some Baumslag--Solitar groups) is 
used in order to construct \emph{finitely summable} spectral triples. This allows to bypass an obstruction 
to finite-summability appearing when using the common regular representation. 
\end{abstract}

Pacs numbers: 11.10.Nx, 02.30.Sa, 11.15.Kc

\section{Introduction}

Lukierski, Ruegg, Nowicki \& Tolstoy discovered a Hopf algebraic deformation of the Poincar\'e Lie algebra and 
called $\kappa $ the deformation parameter \cite{kappafirst1,kappafirst2}. Since this pioneering work, the 
subject became very active: the Hopf algebra was represented on the $\kappa $-deformation of Minkowski 
space \cite{kappaMink1,kappaMink2}. This has been used to generalize the notion of a quantum particle 
\cite{part} or in quantum fields \cite{fields}. Algebraic properties like differential calculi on the 
$\kappa $-Minkowski space were investigated \cite{calculi} as well as the Noether theorem 
\cite{noether,{AmelinoCamelia:2007rn}}. The $\kappa $-Minkowski space has also been popularized as `double
special relativity' \cite{double} and appears in spin foam models \cite{foam}.

It is natural to ask whether $\kappa $-geometry is a noncommutative one in the sense 
of Connes \cite{triple,triple1} (see \cite{answer} for a first attempt). While the algebraic setting is quite clear, the 
main difficulty is to overcome the analysis which is an essential part in the definition of spectral triples.

The $\kappa$-deformation of $n$-dimensional Minkowski space is based on the Lie-algebraic relations
\begin{align}
\label{commutationkappa}
[x^0,\,x^j]:= \tfrac{i}{\kappa} \,x^j, \quad [x^j,\,x^k]=0, \quad j,k=1,\dots,n-1.
\end{align}
Here we assume $\kappa>0$. As in \cite[eq. (2.6)]{KMLS}, one gets 
\begin{align*}
e^{i c_\mu x^\mu}=e^{i c_0 x^0}\, e^{i c'_j \,x^j} \text{ where } 
c'_j:= \tfrac{\kappa}{c_0}\,(1-e^{-c_0/\kappa}) c_j.
\end{align*}
Assuming that the $x^\mu$'s are selfadjoint operators on some Hilbert space, we define unitaries  
$$
U_\omega :=e^{i\omega x^0} \text{ and }V_{\vec{k}}:=e^{-i\sum_{j=1}^{n-1} k_jx^j}
$$
with $\omega,k_j \in \R$, which generate the $\kappa$-Minkowski group considered in \cite{Agostini}.

 If $W(\vec{k},\omega):=V_{\vec{k}}\,U_\omega$, one gets as in \cite[eq. (13)]{Agostini}
\begin{align}
\label{grouplaw}
 W(\vec{k},\omega) \, W( \vec{k'},\omega')=W(e^{-\omega/\kappa} \vec{k'}+\vec{k},\omega + \omega').
\end{align}
The group law \eqref{grouplaw} is, for $n=2$, nothing else but the crossed product 
\begin{align}
\label{groupkappa}
G_\kappa:=\R \rtimes_\a \R \text{ with group isomorphism }\,\a(\omega)k:=e^{-\omega/\kappa}k, \,\,\, k\in \R.
\end{align}
Note that $G_\kappa \simeq \R\rtimes \R^*_+$ is the affine group on the real line which is solvable and 
nonunimodular. The irreducible unitary representations are either one-dimensional, or fall into two 
nonequivalent classes \cite{Agostini}. When $\kappa \rightarrow \infty$, the usual plane $\R^2$ is recovered but 
with an unpleasant pathology at the origin \cite{DP}.

For a given $\omega$, a particular case occurs when $m :=e^{-\omega /\kappa}  \in \N^*$, since 
\begin{align}
\label{commutationgen}
 U_\omega V_{\vec{k}}=(V_{\vec{k}})^m U_\omega,
 \end{align}
 $m$ being independent of $k_j$. This means that for chosen $\omega$ and $k_j$, the presentation of the group 
 is given by two generators and one relation.

Here, we investigate different spectral triples ($\A,\H,\DD)$ associated to the group 
$C^*$-algebra of $G_\kappa$.  To avoid technicalities due to a continuous spectrum of $\DD$, we want a 
unital algebra so that we consider a periodic time $x^0$ which induces a discrete version $G_a$ of $G_\kappa$ 
where $a$ is a real parameter depending on $\kappa$. This is done in section \ref{motivation}.
 
In section \ref{aquelconque}, we give the main properties of the algebra $\U_a = C^*(G_a)$ and its 
representations. 
For $C^*$-algebras of groups, the left regular representation is the natural one to consider. But due to the 
structure of $G_a$ (solvable with exponential growth, see Theorem~\ref{Thealgebra}), there is a known 
obstruction to construct finite-summable spectral triples on $\U_a$ based on this representation. In order to 
bypass this obstruction,  
we need to refine our understanding of the structure of $G_a$ in terms of an underlying dynamical system. This 
structure permits to define a particular representation of $\U_a$ using only the 
periodic points of the dynamical system. Following Brenken and J\o rgensen \cite{BJ}, the 
topological entropy of this dynamics is considered. 

It is worthwhile to notice that the elementary building blocks $G_a$ 
given by $a=m \in \N^*$ as in \eqref{commutationgen} are some of the amenable  Baumslag--Solitar groups, 
already encountered in wavelet theory \cite{J}. The power of harmonic analysis on groups also justifies a 
reminder of their main properties in section \ref{abstractapproach}. 

The question of the finite summability of these 
triples is carefully considered with results by Connes \cite{Con89} and Voiculescu 
\cite{Voiculescu79, Voiculescu}: 
using the regular representation of $C^*(G_a)$, an obstruction to finite-summability appears and allows only 
$\theta$-summability.
However, faithful representations of $C^*(G_a)$, not quasi-equivalent to the left regular one and based on the 
existence of periodic points for dynamical systems, can give rise to arbitrary finite-summable spectral triples. 
These results are summarized in Theorems \ref{nonexistence} and \ref{existence}. 
\\ All proofs will appear elsewhere.

\section{Motivations and models}
\label{motivation}

Let us consider the example given by \eqref{commutationkappa} for two hermitian generators $x^0$ 
and $x^1$. For any $(k,\omega) \in \R^2$, one defines 
$W(k,\omega) := V_{k} \,U_\omega = e^{-ik x^1} e^{i \omega x^0}$. Then one has \eqref{grouplaw} which is
a representation of the group $G_\kappa$ defined in \eqref{groupkappa}. The bounded operators
$W(f) := \int_{G_\kappa} f(k, \omega) \,W(k, \omega) \, e^{\omega/\kappa}  \,dk \,d\omega$
for any $f \in L^1(G_\kappa, e^{\omega/\kappa}  \, dk\,d\omega)$ (here $e^{\omega/\kappa} dk\,d\omega$ is 
the left Haar measure on $G_\kappa$) generate a representation of $C^*_{red}(G_\kappa)$. 
The product of $f, g \in L^1(G_\kappa, e^{\omega/\kappa}  \, dk\,d\omega)$ takes the 
form $(f \ast_\kappa g)(k, \omega) = \int_{G_\kappa} f(k', \omega') g(e^{\omega'/\kappa}(k - k'), \omega - \omega') 
\,e^{\omega'/\kappa} \,dk'\,d\omega'$. 
The advantage of considering the theory of group $C^*$-algebras is twofold. Many structural 
properties on groups will turn out to be useful in studying some properties of the corresponding $C^*$-algebras. 
Moreover, this allows us to construct in a natural way compact versions of noncommutative spaces as we 
now explain.

For an abelian topological group $G$, $C^*_{red}(G)$ is isomorphic to $C_0(\widehat{G})$ where 
$\widehat{G}$ is the Pontryagin dual of $G$. Both algebras are defined as spaces of functions. By duality, a 
discrete subgroup $\Gamma \subset G$ produces the $C^*$-algebra 
$C^*_{red}(\Gamma) \simeq C(\widehat{\Gamma})$ where $C(\widehat{\Gamma})$ is the $C^*$-algebra of 
continuous functions on the compact space $\widehat{\Gamma}$. Notice that there is a natural dual map 
$\widehat{G} \rightarrow \widehat{\Gamma}$. For the example of the plane, consider the discrete subgroup 
$\Gamma = \Z^2 \subset \R^2$. Then the resulting $C^*$-algebra is $C^*(\Z^2) \simeq C(\T^2)$ because 
$\widehat{\Z} = \T^1$. The dual map $\R \simeq \widehat{\R} \rightarrow \widehat{\Z} = \T^1$ is explicitly 
given by $x \mapsto e^{2\pi ix}$. The choice of the subgroup $\Gamma = \Z^2 \subset \R^2$ corresponds 
then to the choice of the compact version $\T^2$ of the (dual) space $\widehat{\R}^2 \simeq \R^2$. The 
compactification takes place in the space of the variables $(x,y)$. 

In order to get a compact version of this $\kappa$-deformed Minkowski 
space, one has to choose a discrete subgroup $H_\kappa \subset G_\kappa$. Since $H_\kappa$ is discrete and 
non-abelian, the associated algebra $C^*(H_\kappa)$ is unital and noncommutative, so it can be interpreted as 
a compact noncommutative space. This point is motivated in section \ref{spectraltriples} and is done in section 
\ref{themodel-subgroup}.

As a final preliminary remark, let us mention that the groups we will encounter will be decomposed as crossed 
products with $\Z$, so both the (related) theories of discrete dynamical systems and crossed products of 
$C^*$-algebras will be intensively used in many parts of this work.

\subsection{Spectral triples}
\label{spectraltriples}
The goal is to study the existence of spectral triples for the $\kappa$-deformed space. A 
spectral triple (or unbounded Fredholm module) $(\A,\H,\DD)$ \cite{Con95,triple,ConnesMarcolli} is given by a 
unital $C^*$-algebra $A$ with a faithful representation $\pi$ on a Hilbert space $\H$ and an unbounded
self-adjoint operator $\DD$ on $\H$ such that 

\qquad - the set $\A=\set{a \in \A \,: \, [\DD, \pi(a)] \text{ is bounded }}$ is norm dense in $A$,

\qquad - $(1+\DD^2)^{-1}$ has a compact resolvent.

($\A$ is always a $^*$-subalgebra of $A$.)
\\Of course, the natural choice of the algebra $A$ is to take the $C^*$-algebra of the group $G_\kappa$, but 
since $A=C^*(G_\kappa)$ has no unit, we need to replace the second axiom by:

\qquad - $\pi(a) (1+\DD^2)^{-1}$ has a compact resolvent for any $a\in \A$.
\\This technical new axiom generates 
a lot of analytical complexities but is necessary to capture the metric dimension associated to $\DD$. For 
instance, if a Riemannian spin manifold $M$ is non-compact, the usual Dirac operator $\DD$ has a 
continuous spectrum on $\H=L^2(S)$ where $S$ is the spinor bundle on $M$. Nevertheless, the spectral  
triple $\big(C^\infty(M),  L^2(S),\DD\big)$ has a metric dimension which is equal to the dimension of $M$.
A noncommutative example (the Moyal plane) of that kind has been studied in \cite{GIGSV}.
\\
We try to avoid these difficulties here by using a unital algebra $A$.

\subsection{The compact version model as choice of a discrete subgroup}
\label{themodel-subgroup}

We consider only dimension $n=2$ but the results can be extended to higher 
dimensions thanks to \eqref{commutationgen}. To get a unit, we choose a 
discrete subgroup $H_\kappa$ of $G_\kappa$ such that $1\in C^*(H_\kappa)$.
\\Since we want also to keep separate the role of the variables $x^0$ and $x^1$, we consider 
the subgroup of the form  $H_\kappa =H \rtimes_\a \Z$: we first replace the second $\R$ of $G_\kappa$ in 
\eqref{groupkappa} by the lattice $\Z$ which corresponds to unitary periodic functions of a chosen frequency 
$\omega_0$ (the time $x^0$ is now periodic). So, given $\kappa>0$ and $\omega_0 \in \R$, with 
\begin{align}
\label{defa}
a:=e^{-\omega_0 /\kappa} \in \R^+,
\end{align}
the group $\R\rtimes_{\a_a} \Z$ is a subgroup of $G_\kappa$ where $\a_a(n)$ is the multiplication by 
$a^n$. This subgroup is a non-discrete ``$ax+b$'' group.

Then, we want a group $H$ to be a discrete (now, not necessarily topological) subgroup of the first $\R$ in 
$\R\rtimes_{\a_a} \Z$, which is invariant by the action $\a_a$. Given $k_0 \in \R$, a natural building block 
candidate for a discrete $H$ is given by $H = B_a \cdot k_0 \simeq B_a$ where 
$$
B_a:=\set{\sum_{i, \,{\rm finite}} m_i\, a^{n_i}  \, : \, m_i,n_i \in \Z},
$$
and more generally, one can take $H\simeq \oplus_{k_0} \,B_a$.

The search for a discrete subgroup $H_\kappa$ of $G_\kappa$ such that $1 \in C^*(H_\kappa)$ 
leads to $H_{\kappa, a} :=B_a \rtimes_{\a_a} \Z$ which is isomorphic to a subgroup of $G_\kappa$ 
once $k_0$ is fixed.

This procedure drives us to the analysis of the algebraic nature of $a$.  
For instance, when $a=m \in \N^*$ is an integer,  this group $H_{\kappa, m}$ is well 
known since it is the solvable Baumslag--Solitar group $BS(1,m)=\Z[\tfrac{1}{m}] \rtimes_{\a_m} \Z$ as shown in 
section \ref{abstractapproach} where we get $B_m=B_{1/m}=\Z[1/m]$. A broad family of noncommutative 
spaces appears:
\begin{lemma}
\label{caseinteger}
In two dimensions, there exists a unital subalgebra $C^*(H_{\kappa, m})$ of the $\kappa$-defor\-ma\-tion 
algebra, associated to the subgroup 
$H_{\kappa, m}=\Z[\tfrac{1}{m}] \rtimes_{\a_m} \Z$ of $G_{\kappa, m}$ 
and  which can be seen as generated by two unitaries $U,V$ such that  
$U=U_{{\omega}_0}$, $V=V_{k_0}$ and $UV=V^mU$. Here,  $\kappa :=-\omega_0\,\log^{-1}(m)>0$ for some 
given integer $m> 1$ and some $\omega_0 \in \R^-$, $k_0\in \R$.
\end{lemma}

\section{\texorpdfstring{The algebra $\U_a$ and its representations}{The algebra Ua and its representations}}
\label{aquelconque}

We now compute the $C^*$-algebra $\U_a$ which is our model for a compact 
version of the $2$-dimensional $\kappa$-Minkowski space. The structure of this algebra is described through a 
semi-direct product of two abelian groups, one of which depends explicitly on the real parameter $a >0$. This 
semi-direct structure gives rise to a dynamical system which is heavily used in the following. The classification of 
the algebras $\U_a$ is performed: the $K$-groups are not complete invariants, and we use the entropy 
defined on the underlying dynamical system to complete this classification. Then some representations of 
$\U_a$ are considered. 
They strongly depend on the algebraic or transcendental character of $a$. In the algebraic case, some 
particular finite dimensional representations are introduced based on periodic points of the dynamical system. 
This construction will be used in section~\ref{exist}.

Let $a=e^{-\omega_0/\kappa} \in \R^*_+$ with $a \neq 1$, and let us recall general facts from \cite{BJ}:
\\Define
$$
B_a:=\set{ \sum_i    m_i \,a^{n_i}  \text{ for finitely many } m_i,n_i \in \Z  }.
$$
This discrete group is torsion-free so its Pontryagin dual $\widehat B_a$ is connected and compact.  

Let $\a_a$ be the action of $\Z$ on $b \in B_a$ defined by $\a_a(n)b:=a^n \,b$, let $\widehat{\a_a}$ be the 
associate automorphism on $\widehat B_a$ and
\begin{align*}
G_a & :=B_a \rtimes_{\a_a} \Z,
&
\U_a & := C^*(G_a) = C^*(B_a) \rtimes_{\a_a} \Z = C(\widehat B_a) \rtimes_{\widehat{\a_a}} \Z.
\end{align*}
This kind of $C^*$-algebras also appeared in \cite{CPPR} for totally different purposes. The group $G_a$ is 
generated by $u:=(0,1) \text{ and } v:=(1,0).$

\begin{lemma}
\label{symmetry}
Let $a\in \R_+^*$, then $B_a=B_{1/a}$ and 
$G_a \simeq G_{1/a}$. Thus the $C^*$-algebra $\U_a$ and $\U_{1/a}$ are isomorphic.
\end{lemma}
The symmetry point $a=1/a$ corresponds to the 
commutative case in \eqref{commutationgen} with $a=1$ or the undeformed relation \eqref{commutationkappa} 
with $\kappa=\infty$. In this spirit $\U_a$ can be viewed as a deformation of the two-torus.

The dynamical system $\U_a\simeq C(\widehat B_a) \rtimes_\a \Z$ has an ergodic action (if $a\neq1$)
 \cite{BJ,Brenken} and if the set of $q$-periodic points is 
$$
\mathrm{Per}_q(\widehat B_a) := \set{\chi  \in \widehat B_a \, : \,  \widehat{\a}^{k}(\chi) \neq  \chi, 
\ \forall  k<q, \ \widehat{\a}^{q}( \chi) =  \chi}
$$
then the growth rate 
$\lim_{q\rightarrow \infty} q^{-1}\log \big(\# \mathrm{Per}_q(\widehat B_a)\big)$ of this sets is an invariant of 
$\U_a$ which coincides with the topological entropy $h(\widehat{\a_a})$:

\begin{align}
\label{entrop}
h(\widehat{\a_a})=\lim_{q\rightarrow \infty} q^{-1}\log \big(\# \mathrm{Per}_q(\widehat B_a)\big).
\end{align}
This entropy can be finite or infinite, dividing the algebraic properties of $a$ into two cases: $a$ can be a 
an algebraic or a transcendental number.

\subsection{Transcendental case}
\label{transcendent}

If $a$ is a transcendental number, then $B_a\simeq \Z[a,a^{-1}]$. Thus, $B_a\simeq \oplus_\Z \Z$, 
$\widehat B_a\simeq S_a:= \set{z=(z_k)_{k=-\infty}^\infty  \in \T^\Z}$ and $\big(\widehat{\a} (z)\big)_k=z_{k+1}$ 
for $z\in S_a, \,k\in \Z$ so that $\widehat{\a}$ is just the shift $\sigma$ on $\T^\Z$. Thus 
$\U_a \simeq C(\T^\Z)\rtimes_\sigma \Z \, \text{ and } \,h(\widehat{\a})=\infty.$ 
Note that the wreath product $\wr$ appears with its known presentation:
$$
G_a=B_a \rtimes_\a \Z  \simeq \Z \wr \Z \simeq  \langle u,v \, : \, [u^i v u^{-i},v]=1 \text{ for all } i\geq1 \rangle.
$$
This group is amenable (solvable), torsion-free, finitely generated (but not finitely presented), 
residually finite with exponential growth. 
\\$S_a$ contains a lot of $q$-periodic points: they are obtained by repeating any  
sequence $(z_k)_{k=0}^{q-1}$ of arbitrary elements in $\T$. Aperiodic points are easily constructed also.
Moreover, $S_a$ is the Bohr compactification $b_{B_a}\R$ of $\R$.

\subsection{Algebraic case}
\label{algebrique}

Assume now that $a$ is algebraic. Let $P\in \Q[x]$ be the monic irreducible polynomial such that $P(a)=0$ and 
let $P=c\,Q_a$ where $Q_a\in \Z[x]$. If $d$ is the degree of $Q_a$, we get the ring 
isomorphism  $B_a  \simeq \Z[x,x^{-1}]/(Q_a)$ \cite{BJ}.
Moreover, $B_a$ has a torsion-free rank $d$. 
If $Q_a(x)=\sum_{j=0}^d q_j\, x^j$ 
(so $Q_a$ has leading coefficient $q_d \in \N^*$), let $A_a \in M_{d\times d}(\Z)$ 
be the $d\times d$-matrix defined by $({A_a})_{i,j}:=q_d \,\delta_{i,j-1}$ for $1\leq j\leq d $ and 
$(A_a)_{d,j}=-q_{j-1}$. Then $q_d\,a^j=\sum_{k=1}^d (A_a)_{j,k}\, a^{k-1}$.
\\For instance, if $a=1/m$ for $m\in \N^*$ then $P(x)=x-1/m$, $Q_a(x)=mx-1$, $d=1$, so $B_a=\Z[\frac{1}{m}]$ 
and $A_a$ is just the number $1$.

Let $\sigma$ be the shift on the group ${(\T^d)}^\Z$ and consider its $\sigma$-invariant subgroup 
$K_a:=\set{z=(z_k)_{k=-\infty}^\infty \in {(\T^d)}^\Z \, : \, q_d\,z_{k+1}=A_a \,z_k}$ (use $\T \simeq \R/\Z$). 
\\
If $S_a$ is the connected component of the identity of $K_a$, then there 
exists a topological group isomorphism $\psi \, : \, \widehat B_a \rightarrow S_a$ such that 
$\sigma_{\vert {S_a}} \circ \psi = \psi \circ \widehat{\a}$ \cite[Theorem 19]{Lawton}. For any 
$\chi \in \widehat B_a$, the associated $z=(z_k)_{k=-\infty}^\infty$ is given by $z_k^{(i)} = \chi(a^{k+i-1})$ 
where $z_k = (z_k^{(i)})_{i=1}^{d} \in \T^d$. In particular $z_0$ is given by 
$\big(\chi(1), \chi(a), \dots, \chi(a^{d-1})\big)$. This map is only surjective on the connected 
component of the identity.
\\
When $a=1/m$ with $m\in \N^*$, we will recover $S_{1/m}=S_m$ in \eqref{solenoid}.

There is a morphism of groups $\hat{\iota} : \R^d \rightarrow S_a$ defined as follows: 
to any $\phi = (\phi^{(i)})_{i=1,\dots,d} \in \R^d$, one associates 
$\hat{\iota}(\phi) = z = (z_k)_{k=-\infty}^\infty \in {(\T^d)}^\Z$ with
\begin{equation}
\label{eq-thetamap}
z_k^{(i)} = \exp\big( 2 i \pi q_d^{-k} \sum_{j=1}^{d} (A_a^k)_{i,j} \phi^{(j)} \big).
\end{equation}
This shows that $S_a$ is a Bohr compactification of $\R^d$ \cite[Proposition 2.4]{Brenken}. 
Then $\widetilde{\a}(\phi) = q_d^{-1} A_a \phi$ defines an action of $\Z$ on $\R^d$ which 
satisfies $\hat{\iota} \circ \widetilde{\a} = \widehat{\a} \circ \hat{\iota}$.
\\
If $r_i$, $i=1,\cdots d$ are the roots of $P$, then by \cite[Proposition 3, Corollary 1]{BJ},
\begin{align}
\label{cq}
c_q(a):=\#\text{ Per}_q(S_a)=\Pi_{k=1}^q \vert Q_a(e^{i2\pi k/q})\vert 
= \vert q_d \vert^q \, \Pi_{k=1}^d \vert 1-{r_k}^q \vert.
\end{align} 
Thus by \eqref{entrop}, the topological entropy is
\begin{align}
\label{entropy}
h(\widehat{\a_a})=\log\vert q_d \vert +\sum_{i, \,\vert r_i\vert >1} \log \vert r_i\vert.
\end{align}
In case $a=m$ or $a=1/m$, \eqref{entropy} gives $h(\widehat{\a_m})=h(\widehat{\a_{1/m}})=\log(m)$. 
\\ Aperiodic points in $S_a$ can be easily constructed using the map $\hat{\iota}$ defined by 
\eqref{eq-thetamap}: any $\phi \in (\R\backslash \Q)^d$ defines an aperiodic point $\hat{\iota}(\phi) \in S_a$.

\subsection{\texorpdfstring{On the structure and classification of algebras $\U_a$}{The structure of algebras Ua}}

\noindent We can now give the main properties of the algebras $\U_a$:
\begin{theorem}
\label{Thealgebra}
Let $a \in \R^*_+$ and $a\neq 1$. Then

(i) The group $G_a = B_a \rtimes _\a \Z$ is a torsion-free discrete solvable group with exponential growth and 
$\widehat{B_a}$ is a compact set isomorphic to a solenoid $S_a$.

(ii) $\U_a=C^*_{red}(B_a \rtimes_\a \Z)\simeq C(S_a) \rtimes_{\widehat{\a}} \Z$ is a NGCR
\footnote{A $C^*$-algebra $A$ is said to be CCR or liminal if $\pi(A)$ is equal to the set of compact operators on 
the Hilbert space $\H_\pi$ for every irreducible representation $\pi$. The algebra $A$ is called NGCR if it has no 
nonzero CCR ideals. \\
An $AF$-algebra is a inductive limit of sequences of finite-dimensional $C^*$-algebras. \\
The algebra $A$ is residually finite-dimensional if it has a separating family of finite-dimensional 
representations.}, AF-embeddable, non-simple, residually finite dimensional $C^*$-algebra and its generated 
von Neumann algebra for the left regular representation is a type $\mathrm{II}_1$- factor. 
%
%
\end{theorem}
\noindent A main point of this theorem, crucial for the sequel is that the algebra $\U_a$ is residually finite and its 
proof is based on properties of the underlying dynamical system:
\begin{prop}
\label{prop-dynsyst}
Let $a \in \R^*_+$ and $a\neq 1$. The subgroup of periodic points and the set of aperiodic points of $S_a$ 
under $\widehat{\a}$ are dense. The space of orbits of $S_a$ is not a $T_0$-space.
\end{prop}
\noindent The classification of algebras $\U_a$ is also based on the dynamical system:

\begin{theorem}
\label{classification}
Let $\omega_0 \in \R$ and $\kappa \in \R^*_+$ defining $a\neq 1$ in \eqref{defa}. 

(i) $\U_a \simeq \U_{a'}$ yields $c_q(a)=c_q(a'),\, \forall q\in \N^*$.

(ii) The entropy $h(\widehat{\a})$ is also an isomorphism-invariant of $\U_a$.
\end{theorem}
This result has important physical consequences since a full Lebesgue measure dense set of different 
parameters $a$ (namely the transcendental ones) generates the same algebra or  $\kappa$-deformed space, 
while in the rational case, these spaces are different:

\begin{corollary}
\label{classification1}
As already seen, $\U_a\simeq \U_{1/a}$. Moreover, 

(i) All transcendental numbers $a$ generate isomorphic algebras $ \U_a$.

(ii) If $\U_a\simeq \U_{a'}$, then $a$ and $a'$ are both simultaneously algebraic or transcendental numbers.

(iii) If $\U_a\simeq \U_{a'}$, then $a'=a$ or $a'=a^{-1}$ in the following cases: $a$, $a'$ or their inverses are in 
$\Q^*$ or are quadratic algebraic numbers.

(iv) If $a=m/l \in \Q^*_+$, $K_0(\U_a) \simeq \Z$ and $K_1(\U_a) \simeq \Z \oplus \Z_{l-m}$.
\end{corollary}
Using \cite{Brenken2}, we can show that the $K$-groups do not give a complete 
classification even in this algebraic case.

\subsection{\texorpdfstring{On some representations of $\U_a$ for algebraic $a$}{On some representations 
of Ua for algebraic a}}
\label{representationsinduced}
We will concentrate on $a\neq1$ algebraic and follow the construction of finite dimensional representations 
\cite{ST,Yamashita}: Let $z_q\in \mathrm{Per}_q(S_a)$ be a $q$-periodic point of $\widehat{\a}$. Let 
$\rho_{z_q}: C(S_a) \rightarrow M_{q}(\C)$ be a representation of $C(S_a)$ defined by
$$
\rho_{z_q}(f):=\text{Diag} \big( \, f(z_q),\cdots,f\big(\widehat{\a}^{q-1}(z_q)\big) \,\big) \in M_{q}(\C)
$$
and for $x\in \T$, let $u_{x,z_q}:= \genfrac{(}{)}{0pt}{1}{\,0 \quad x}{1_{q-1}\,\,\,\, 0 } \in M_q(\C)$. 
It is a unitary which satisfies the covariance relation 
$u_{x,z_q}^* \, \rho_{z_q}(f) \,u_{x,z_q}=\rho_{z_q}(f \circ \widehat{\a})$
thus $\pi_{x,z_q}:=\rho_{z_q} \rtimes u_{x,z_q}$ is a representation of $\U_a$ on $M_q(\C)$. Again, 
$\pi_x:=\oplus_{q=1}^\infty \oplus_{z_q \in \mathrm{Per}_q (S_a)}$ is a representation of $\U_a$. 
So, for a dense family $\set{x_l}_{l=1}^\infty$ in $\T$, using the canonical faithful conditional expectation 
$C(S_a) \rtimes_{\widehat{\a}} \Z \rightarrow C(S_a)$ and the density of periodic points, one show that 
$\pi:=\oplus_{l=1}^\infty \pi_{x_l}$ is a faithful representation of $\U_a$ such that 
$\pi(\U_a) \subset \oplus_{l=1}^\infty \oplus _{q=1}^\infty \oplus _{z \in \mathrm{Per}_{q}(S_a)} M_{q}(\C)$.

The representation $\pi_{x,\chi}$ where $\chi=z_q \in \widehat{B_a}=S_a$, can be extended to a representation 
$\pi_\chi$ on the Hilbert space $\H_\chi = L^2(\T) \otimes \C^q$ by 
\begin{equation}
\label{eq-reppichi}
\pi_\chi = \int_{\T} \pi_{x,\chi}\, dx.
\end{equation}
Denote by $\{e^{(q)}_s \}_{s=1, \dots q}$ the canonical basis of $\C^q$ and by 
$e_n : \theta \mapsto e^{in\theta}$, for $n \in \Z$, the basis of 
$L^2(\T) \simeq \ell^2(\Z)$. Then let, for $f \in C^\ast(B_a) = C(\widehat B_a)$:
\begin{align*}
\pi_\chi(f) (e_n \otimes e^{(q)}_s) &:= f \circ \widehat{\a}^{s-1}(\chi) \, e_n \otimes e^{(q)}_s, \\
U (e_n \otimes e^{(q)}_s) &:= 
\begin{cases}
e_n \otimes e^{(q)}_s & \text{for $1\leq s < q$},\\
e_{n+1} \otimes e^{(q)}_1 & \text{for $s = q$},
\end{cases}
\end{align*}
where $U$ is the generator of $\Z$. This representation is constructed from the representation of 
$G_a=B_a \rtimes_\a \Z$ given by 
$\pi_\chi(b) (e_n \otimes e^{(q)}_s) = \chi \circ \a^{s-1}(b) \, e_n \otimes e^{(q)}_s$ for any $b \in B_a$ (the 
generator $U$ of the action of $\Z$ is the same).

Another natural representation to consider is the representation of $\U_a$ obtained from the left regular 
representation of $G_a$.

In \cite{LPT} and more systematically in \cite{DJ}, the induced representations \textit{\`a la} Mackey of $G_a$ for 
algebraic $a$ have been investigated. The main results are the following.

For any $\chi \in \widehat B_a$, let the space of functions 
$\varphi : B_a \rtimes_\a \Z \rightarrow \C$ such that $\varphi(b,k) = \chi(b) \varphi(0,k)$ for any $b \in B_a$ and 
$k \in \Z$ be endowed with the norm $\| \varphi \|_\chi^2 = \sum_{k \in \Z} |\varphi(0,k)|^2$. This defines a Hilbert 
space denoted by $\H_\chi^{\mathrm{Ind}}$. The induced representation of $G_a$ on 
$\H_\chi^{\mathrm{Ind}}$ is given by $(\pi_\chi^{\mathrm{Ind}}(g) \varphi)(h) = \varphi(hg)$ for any $g,h \in G_a$. 

This representation is unitarily equivalent to the following $\pi'_\chi$ \cite[Theorem~4.2]{DJ}: the 
Hilbert space is $\ell^2(\Z)$ and for any $\xi = (\xi_k)_{k \in \Z}$, one takes 
$(\pi'_\chi(b)\xi)_k := \chi\circ \a^{k}(b) \xi_k$ and the generator of $\Z$ is $(U \xi)_k = \xi_{k+1}$. As a 
representation of $\U_a$, one has $(\pi'_\chi(f)\xi)_k = f \circ \widehat{\a}^{k}(\chi) \xi_k$ for any 
$f \in C^\ast(B_a)$.

\begin{theorem}
\label{thm-inducedrepresentations}
Assume $a\neq 1$ is algebraic.

(i) There is a natural bijection between the set of orbits of $\widehat{\a}$ in $S_a$ and the set of all equivalence 
classes of induced representations of $\U_a=C^\ast(G_a)$. This bijection is realized by 
$\chi \mapsto \pi_\chi^{\mathrm{Ind}}$.

(ii) The representation $\pi_\chi^{\mathrm{Ind}}$ is irreducible if and only if $\chi$ is aperiodic.

(iii) The commutant of $\pi_\chi^{\mathrm{Ind}}$ for a $q$-periodic point $\chi$ is the commutative algebra 
$C(\T)$.

(iv) The right regular representation $R$ of $\U_a=C^\ast(G_a)$ is unitarily equivalent to the representation
$\int_{B_a}^{\oplus} \pi_\chi^{\mathrm{Ind}} \, d\mu(\chi)$.
\end{theorem}
\noindent For a $q$-periodic $\chi$, the representation $\pi_\chi^{\mathrm{Ind}}$ is reducible. Explicitly one has:

\begin{prop}
If $\chi$ is $q$-periodic, $\pi_\chi^{\mathrm{Ind}}$ is unitarily equivalent to the representation 
$\pi_\chi$ on $\H_\chi$, so that its continuous decomposition into irreducible finite dimensional representations 
on $\C^q$ is realized by \eqref{eq-reppichi} along $\T$.
\end{prop}
This proposition states that, while the finite dimensional representations of $\U_a$ are not obtained as induced 
representations, they are nevertheless reductions of induced representations. The right regular representation 
$R$ contains the infinite dimensional irreducible induced representations which are only accessible using 
aperiodic points. The representations $R$ and $\pi_\chi$ are not quasi-equivalent: this difference will play a 
crucial role in the construction of different spectral triples, see Remark \ref{Contradict}.

While the $\pi_\chi^{\mathrm{Ind}}$'s yield a von Neumann factor of type $\mathrm{I}$, $R$ gives 
a type $\mathrm{II}_1$ factor because of the integral. So the group $G_a$ is non-type $\mathrm{I}$.

\section{\texorpdfstring{The particular case $a=m \in \N^*$}{The particular case a=m in N*}}
\label{abstractapproach}

According to Lemma \ref{symmetry}, the case $a=m \in \N^*$ also covers the case 
$a=1/m$. We do insist on this $\kappa$-deformed space since the algebra 
is then generated by two unitaries related by one relation (see \eqref{def} below) in the spirit of the  
noncommutative two-torus: $G_m=BS(1,m)$ is the Baumslag--Solitar group which is generated by two 
elements and one-relator while, when $a$ and $a^{-1}$ are not integers, $G_a$ is not a finitely presented group 
(still with two generators). This simplifies the computations of section \ref{algebrique}.

Moreover, the results described now rely more on some properties of the 
Baumslag--Solitar group than on the dynamical system. Thus, these results 
(which for the most are already valid and exposed for generic values of $a$) are presented 
independently. These structures appear also naturally in wavelet theory, which could benefit from our analysis.

\subsection{The algebra}
\label{thealgebra}

\begin{definition}
\label{def-thealgebra}
Let $\U_m$ be the universal $C^*$-algebra labelled by $m \in \Z^*$ (restricted to $\N^*$ later) and 
generated by two unitaries 
$U$ and $V$ such that  
\begin{align}
\label{def}
UVU^{-1}=V^m.
\end{align}
\end{definition}
This universal $C^*$-algebra $\U_m$ is denoted by $\OO(E_{1,m})$ in \cite{Kat}, and also $\OO_{m,1}(\T)$ in 
\cite{Yamashita} (where only $m\in\N^*$ is considered.) These algebras are topological graph $C^*$-algebras 
which can be seen as transformation group $C^*$-algebras on solenoid groups as already noticed in 
\cite{BJ,J,Brenken}. They have been used in wavelets and coding theory \cite{DHPQ,DJ,DJP}. 

Relation \eqref{def} also appeared in the Baumslag--Solitar group $BS(1,m)$ introduced in \cite{BS} as the 
group generated by $u,\,v$ with a one-relator:
$$
BS(1,m):=\langle u,\,v \,\vert \, uvu^{-1}=v^m \rangle.
$$
This group plays a role in combinatorial and geometric group theory. It is a finitely generated, meta-abelian, 
residually finite, Hopfian, torsion-free, amenable (solvable non-nilpotent) group. It has infinite conjugacy classes,  
a uniformly exponential growth (for $m\neq 1$) but is not Gromov hyperbolic \cite{Harpe}. 
Note that $BS(1,1)$ is the free abelian group on two generators and $BS(1,-1)$ is the Klein bottle 
group. As for the $BS(1,m)$ groups, within the algebras $\U_m$, we remark that $\U_1$ and $\U_{-1}$ play a 
particular role: $\U{_1} = C(\T^2)$ and  
$\U_{-1}\supset C(\T^2)$ will not be considered here since we need $a=m>0$.

For $m\geq 2$, a solenoid appears as in section~\ref{algebrique}, as well as a crossed product structure, a fact 
that we recall now in this particular context.

Assume $2\leq m \in \N$ and let the subring of $\Q$ generated over $\Z$ by $\tfrac{1}{m}$
$$
B_m=B_{1/m}:=\Z [\tfrac{1}{m}]:=\bigcup_{l\in \N}m^{-l} \Z \subset \Q.
$$
It is the additive subgroup of $\Q$ which is an inductive limit of the rank-one groups $m^{-l}\Z$, for 
$l=0,1,2,\dots$ and $B_m$ has a natural automorphism $\a$ defined by 
$\a(b):=mb$. 
Note that the abelian group $B_m$ is not finitely generated. When $m\rightarrow \infty$, 
$BS(1,m) \rightarrow \Z \wr \Z $ (in the space of marked groups on two generators) \cite{Stalder}. This group also 
appears when $m=e^{-\omega_0/\kappa}$ is replaced by a transcendental number $a\in \R^*_+$ as seen in 
section \ref{aquelconque}.

$B_m$ can be identified with the subgroup of the affine group Aff$_1(\Q)$ generated by the dilatation 
$u:x \rightarrow mx$ and the translation $v: x\rightarrow x+1$. It is the subgroup normally generated in 
$BS(1,m)$ by $\langle v, u^{-1} vu, u^{-2} v u^2 , . . .\rangle$.
The Baumslag--Solitar group $BS(1,m)$ is then isomorphic to the crossed product 
$BS(1,m) \simeq B_m \rtimes_\a \Z$ so that one has the group extension 
$$
1\rightarrow  \Z[ \tfrac{1}{m}] \rightarrow B(1, m) \rightarrow \Z \rightarrow 1. 
$$
\noindent Using this crossed product decomposition, the group $BS(1,m)$ has the following explicit law: 
$(b,l)(b',l')=(b+\a^l(b'),l+l')$ for $l,l'\in \Z$ and $b,b'\in B_m$. It is of course generated by the elements $u:=(0,1)$ 
and $f_b:=(b,0)$ with $b\in B_m$. Thus for $j,l\in \Z, n\in\N$,  $uf_bu^{-1}=f_{\a(b)}$, and
\begin{align}
\label{gen}
\text{if }f_{\a^{-n}}j:=(\a^{-n}j,0)=u^{-n}f_j u^n, 
\text{ then } ((\tfrac{1}{m})^{n}j,l)=f_{{(\tfrac{1}{m})}^nj}\,u^l.
\end{align} 

$BS(1,m)$ is a subgroup of the $``ax+b"$ group (endowed with the law 
$(b,a)(b',a'):=(b+ab',aa')$) and can be viewed as the following subgroup of two-by-two matrices 
$\set{ \genfrac{(}{)}{0pt}{1}{\,m^l \,\,\, b\,}{\,\,0 \,\,\,\,\,\,1\, }  \; : \; l\in \Z,\,b\in B_m}$.  
$BS(1,m)\simeq B(1,m')$ is equivalent to $m=m'$ \cite{Mol}.

Let $\widehat {B_m}$ be the Pontryagin dual of $B_m$ endowed with the discrete topology. It is isomorphic to 
the solenoid
\begin{align}
\label{solenoid}
S_m =S_{1/m}\simeq \set{(z_k)_{k=0}^{\infty} \in \prod_{i=0}^\infty \T \, : \, z_{k+1}^m=z_k, \,\forall k\in \N_0}
\end{align}
using $z_k:=\chi\big((\tfrac{1}{m})^k\big)$ for any $\chi \in \widehat B_m$. The group $S_m$ is compact 
connected and abelian. Notice that (see section~\ref{algebrique})\\ 
$S_m \simeq \set{(z_k)_{k=-\infty}^{\infty} \in  \T^\Z \, : \, z_{k+1}^m=z_k, \, k\in \Z}$ 
defining $z_{-k}:=z_0^{mk}$ for $k>0$.

The embedding 
$\hat{\iota} : \theta \in\R \mapsto \chi_\theta \in S_m \text{ where }\chi_\theta(b):=e^{i2\pi \theta b} \in \T$ for  
$b\in B_m$ identifies $S_m$ as the Bohr compactification $b_{B_m}\R$ of $\R$.

$S_m$ is endowed with a natural group automorphism $\widehat\a$ given by 
\begin{align*}
\widehat{\a} (z_0,z_1,z_2,\dots)=(z_0^m,z_0,z_1,\dots) \qquad
\widehat{\a} ^{-1}(z_0,z_1,z_2,\dots)=(z_1,z_2,\dots).
\end{align*}

All $q$-periodic points in $S_m$ are of the following form: if $z_0$ is a solution of $z^{m^{q}-1} =1$, then 
$(z_0,z_0^{m^{q-1}},\dots,z_0^m,z_0,\dots) \in S_m$; so there are only finitely many periodic points, namely 
$c_q(m) = m^{q}-1$ such points.

The $C^*$-algebra $C(S_m) \simeq C^*(B_m)$ is precisely the algebra of almost periodic 
functions on $\R$, with frequencies in $B_m$ and the isomorphism is the map 
$f \mapsto f \circ \hat\iota$. Thus 
$$
\U_m = C^*(BS(1,m)) \simeq C^*(B_m)\rtimes_{\a} \Z \simeq C(S_m) \rtimes_{\widehat\a} \Z.
$$
The unitary element $U$ of Definition~\ref{def-thealgebra} is precisely the generator of the action $\a$ of $\Z$ on 
$C^*(B_m)$ while $V$ is one of the generators $\set{ U^{-\ell}V U^\ell : \ell \in \Z}$ of the abelian 
algebra $C^*(B_m)$. As a continuous function on $S_m$, $U^{-\ell}V U^\ell$ is the function 
$(z_k)_{k=0}^\infty \mapsto z_\ell$ and in particular, $V:(z_k)_{k=0}^\infty \mapsto z_0$.

The subgroup $\set{z:=(z_k)_{k=0}^\infty \in S_m \, : \, \widehat\a^q (z)=z \text{ for some } 
q\in \N^*}$ of periodic points is dense in $S_m$ and $\widehat\a$ is ergodic on $S_m$ for $m\geq 2$ as 
previously seen \cite[Proposition 1]{BJ}.

\subsection{The representations}

The knowledge of $^\ast$-representations of $\U_m$ is essential in the context of spectral 
triples (see Definition \ref{defi}). According to \eqref{gen}, any unitary representation of $BS(1,m)$ is given by a 
unitary operator $U$ and a family of unitaries $T_k$, $k\in \Z$, with the constraint $UT_kU^{-1}=T_{mk}$, so 
there is a bijection between the $^\ast$-representations of $\U_m$ on some Hilbert space $\H$ and the 
corresponding unitary representations of $BS(1,m)$. This is rephrased usefully in the following lemma \cite{J}:

\begin{lemma}
\label{useful}
The algebra $\U_m$ is the $C^*$-algebra generated by $L^\infty(\T)$ and a unitary symbol 
$\tilde U$ with commutation relations, where $e_n(z):=z^n$
\begin{align}
\tilde U \,f\, \tilde U^{-1}=f \circ e_m, \,\,\forall f \in L^{\infty} (\T).
\end{align}
\end{lemma}
$\U_m$ contains a family of abelian subalgebras 
$\A_n:=\tilde U^{-n}\, L^\infty(\T)\,\tilde U^n$ for $n\in \N$, which is increasing since 
$\tilde U^{-n} \,f \,\tilde U^n=\tilde U^{-(n+1)} \,f\circ e_m \, \tilde U^{(n+1)}$.

If we choose the Hilbert space $\H:=L^2(\R)$, the scaling and shift operators give rise to a 
representation $\pi$ of $ \U_m$ on $\H$ by $\pi(U): \psi(x) \mapsto \tfrac{1}{\sqrt{m}} \, 
\psi(\tfrac{x}{m})$ and $\pi(V): \psi(x) \mapsto  \psi(x-1)$. 

The Haar measure $\nu$ on $\widehat B_m$ gives rise to a faithful trace on $\U_m$ and, since there are many 
finite dimensional representations of $\U_m$ (see proof of Theorem \ref{Thealgebra}), there are many traces 
on it.

With $\H:=L^2(S_m,\nu)\simeq \ell^2(B_m)$ and $U: \psi \in \H \mapsto \psi \circ \widehat\a \in \H$, $C(S_m)$ 
acts on $\H$ by left multiplication, we get a covariant representation for $(S_m,U)$ of the dynamical system 
$(C(S_m),\widehat{\a},\Z)$, so a representation of $\U_m$ on $\H$. Since $\widehat\a$ is ergodic, this 
representation is irreducible and faithful \cite[Theorem 1]{BJ}.

If we choose $\H:=\ell^2(\Z)$, for each $\theta \in \R$ we get an induced representation of 
$\U_m$ by $\pi_\theta(U)\psi(k):=\psi(k-1)$ and $\pi_\theta(V)\psi(k):=\chi_\theta( m^{-k})\,\psi(k), k\in \Z$.

\section{On the existence of spectral triples}
\label{exist}

Since we want to construct spectral triples on $\U_a$, it is worthwhile to know the heat decay of 
$G_a=B_a\rtimes_\a \Z$ via a random walk on the Cayley graph of $G_a$ with generators 
$S=\set{x,x^{-1},y,y^{-1}}$ where $x=(0,1)$ and $y=(1,0)$, and with a constant weight and standard Laplacian. 

The decay of the heat kernel $p_t$, with $t\in \N$, has been computed on the diagonal in 
\cite[Theorem 1.1]{Pittet}, \cite[Theorem 5.2]{CGP}: when $t \rightarrow \infty$, we get
$p_{2t} \sim e^{-t^{1/3}\,(\log t)^{2/3}}$ if $a$ is transcendental while $p_{2t} \sim e^{-t^{1/3}}$ if $a$ is 
algebraic. This is related to the fact that $G_a$ has exponential volume growth.

However, for a finite dimensional connected non-compact Lie group, the behaviour of the heat kernel $p_t$ 
depends on $t\in\R^*_+$ and can diverge for the short time behaviour when $t\rightarrow 0$. 
Let us explain how this point is related to the dimension:
In noncommutative geometry, a (regular simple) spectral triple $(\A,\H,\DD)$ has a (spectral) dimension 
which is given by 
$\max \set{n\in \N \, : \, n \text{ is a pole of } \zeta_\DD:s \in \C \rightarrow \Tr(\vert \DD \vert ^{-s})}$ (here 
$\DD$ is assumed invertible). 
In particular, when $M$ is a $n$-dimensional compact Riemannian spin manifold, and $\A=C^\infty(M)$, 
$\H=L^2(S)$ where $S$ is the spinor bundle and $\DD$ is the canonical Dirac operator, the spectral 
dimension coincides with $n$. Via the Wodzicki residue, an integral 
$\ncint X:=\Res_{s=0} \Tr(X \vert \DD \vert^{-s})$ is defined on (classical) pseudodifferential operators $X$ 
acting on smooth sections of $S$. 
For instance, $\ncint \vert \DD \vert^{-n}$ coincides (up to a universal constant) with the Dixmier trace 
$\Tr_{\text{Dix}}(\vert \DD \vert ^{-n})=\lim_{N\rightarrow \infty} \log(N)^{-1}\sum_{k=1}^N \vert \lambda_k 
\vert ^{-n}$ where the $\lambda_k$ are the singular values of $\DD$. The dimension of $M$ appears in 
$\Tr(e^{-t\DD^2}) \sim \sum_{N \geq 0}\tfrac{1}{t^{(n-N)/2}} \,a_N(\DD)$ when $t \rightarrow 0$. In particular, 
when $M=\R^n$ with Lebesgue measure and $\DD^2=-\triangle$ is the standard Laplacian 
(non-compactness is not a problem), the heat kernel is $p_t(x,x)=\tfrac{1}{(4 \pi t)^{n/2}}$ for all $x \in M$ 
(see \cite{Con94,ConnesMarcolli,Polaris}). As a consequence, $\Tr(\vert \DD \vert^{-(n+\epsilon)}) < \infty$ 
for all $\epsilon >0$. 

We will see that, depending on the chosen representation of $\U_a$, such an $n$ does not always 
exist, meaning that the ``dimension is infinite".

\begin{definition}
\label{defi}
A spectral triple (also called unbounded Fredholm module) $(\A,\H,\DD)$ is given by a 
unital $C^*$-algebra $A$ with a faithful representation $\pi$ on a Hilbert space $\H$ and an unbounded
self-adjoint operator $\DD$ such that 

\qquad - the set $\A=\set{a \in \A \,: \, [\DD, \pi(a)] \text{ is bounded }}$ is norm dense in $A$,

\qquad - $(1+\DD^2)^{-1} \in J$ where $J$ is a symmetrically-normed ideal of the compact operators 
$\K(\H)$ on $\H$.

The triple is $p$-summable if $J=\L^p(\H)$ for $1 \leq p<\infty$ which means 
$\Tr\big((1+\DD^2)^{-p/2} \big) <\infty$. It is $p^+$-summable if $J=\L^{p+}(\H)$.

It is finitely summable if it is $p$-summable for some $p$.

It is $\theta$-summable if there exists $t_0\geq 0$ such that $\Tr\big(e^{-t\DD^2} \big) <\infty$ for all $t>t_0$  
(thus $J=\K(H)$).
\end{definition}
\noindent Note that $\A$ is a $^*$-subalgebra of $A$ and $p$-summability implies $\theta$-summability.

Connes proved in \cite{Con89} that, for an infinite, discrete, non-amenable group $G$, there exist no 
finitely summable spectral triples on $A=C^*_{red}(G)$. However,  in this case, there always exist 
$\theta$-summable spectral triples on $A$ (even with $\DD>0$). 
Using a computable obstruction to the existence of quasicentral approximate units relative to $J$ for $A$, 
Voiculescu was able to derive, for solvable groups with exponential growth, the non-existence result for 
unbounded (generalized) Fredholm modules using the Macaev ideal $J=\L^{\infty,1}(\H)$ \cite{Voiculescu}. 
We use these results in the following:

\begin{theorem} 
\label{nonexistence}
Non-existence of finite-summable spectral triples.
\\ Let $ A=\U_a$, $G_a=B_a \rtimes_\a \Z$ and $\A=\C [G_a]$.

(i) There is no finitely summable spectral triple $\big(\pi(\A), \H_\pi,\DD\big)$ when the representation $\pi$ 
is quasiequivalent to the left regular one.

(ii) There exist $\theta$-summable spectral triples $\big(\pi(\A), \H_\pi,\DD\big)$ with $t_0=0$ where the 
representation $\pi$ is quasiequivalent to the left regular one.
\end{theorem}
Despite the previous result, we add a few explicit examples of spectral triples using the fact that the 
algebra $\U_a$ is residually finite. Clearly, these triples deal with a restrictive part of the geometry of the 
$\kappa$-deformation based on $\U_a$, namely the dynamical system which is behind. The residually 
finite property is seen via the periodic points of this dynamics.

\begin{theorem} 
\label{existence}
Existence of finite-summable spectral triples.
\\ Let $ A=\U_a$ and $\A=\C [G_a]$.

(i) There exist spectral triples $\big(\pi(\A), \H_\pi,\DD\big)$ which are compact, i.e. $[\DD,\pi(x)]$ is compact 
for all $x \in \A$.

(ii) There exist spectral triples $\big(\pi(\A), \H_\pi,\DD\big)$ such that $[\DD,\pi(x)]=0$, $\forall x \in \U_a$, 
and with arbitrary summability.

(iii) When $a$ is algebraic, there exist spectral triples $\big(\pi(\A), \H_\pi,\DD\big)$ such that 
$[\DD,\pi(v)]=0$, $[\DD,\pi(u)] \neq 0$ and with an arbitrary summability $p\geq 2$.
\end{theorem}

In case $(i)$, $[\DD,\pi(x)]$ is not necessarily zero but the summability is not controlled while for case $(ii)$, 
the condition $[\DD,\pi(x)]=0$ enables us to control summability. In a sense, case $(iii)$ is a mixed situation 
requiring that $a$ be algebraic. In that situation, we have an explicit representation $\pi$ so 
that formulae for Dirac operators can be proposed.

\begin{remark}
\label{Contradict}
There is no contradiction between Theorems~\ref{nonexistence} and \ref{existence} since the faithful 
quasidiagonal representation (or residually finite one) $\pi$ of $\U_a$ used above to construct $\DD$ is not 
quasiequivalent to the left regular one: actually, as already mentioned, the von Neumann algebra generated 
by $\pi(\U_a)$ is a $\mathrm{II}_1$ factor when $\pi$ is the left regular representation, while it is of type 
$\mathrm{I}$ when $\pi$ is the quasidiagonal or residually finite one \cite[5.4.3.]{Dixmier}. 

A more direct way to confirm that the representation $\pi$ used in the proof of point $(iii)$ of 
Theorems~\ref{existence} is not quasiequivalent to the left regular representation (or to the right regular 
representation which is unitarily equivalent to the left one) is to notice that $\pi$ is the direct integral 
$\pi = \int_{\mathrm{Per}}^{\oplus} \pi_\chi \, d\mu(\chi)$ 
of the \emph{finite dimensional} representations $\pi_\chi$ defined in \eqref{eq-reppichi}. As such, this 
representation is strictly contained in the right regular representation $R$ as can be checked using $(iv)$ of 
Theorem~\ref{thm-inducedrepresentations}. The part of $R$ which is not in $\pi$ is given by the induced 
\emph{infinite dimensional} irreducible representations constructed on aperiodic $\chi$'s.

As noticed in \cite{SZ}, if $(\A,\H_\pi,\DD)$ is a spectral triple with $[\DD,\pi(x)]=0$, $\forall x \in A$, 
then A is a residually finite $C^*$-algebra.
\end{remark}

\begin{remark}
Theorem \ref{nonexistence} says that the 2-dimensional $\kappa$-deformed space reflected by 
the algebra $\U_a$ with $\kappa=-\omega_0\,\log^{-1}(a)$ is in fact ``infinite dimensional" as a metric
noncommutative space. Theorem \ref{existence} is a tentative to restore a metric. For instance, the 
distances on the state space $\SS(\U_a)$ generated by Connes' formula
$$
d(\omega,\omega'):=\sup \set{\vert \omega(a)-\omega'(a)\vert \, :\, a\in \A ,\,  \vert \vert [\DD,a] \vert \vert \leq 1} 
,\quad \omega,\omega' \in \SS(\U_a)
$$
are infinite in case $(ii)\,$of Theorem \ref{existence}, while in case $(iii)\,$some states can be at finite distances.
\end{remark}

\begin{remark}
The operator $\DD$ given in Theorem \ref{existence} $(iii)$ is not directly related to the group structure of 
$G_a$ but rather connected to the underlying dynamical system associated to the algebraic nature of $a$: it 
depends explicitly of the isomorphism-invariant $\set{c_q(a) \, : \,q\in \N^*}$.
\end{remark}

\section{Conclusion}

We have shown that $\kappa$-Minkowski space defined by \eqref{commutationkappa} can be reduced to a 
compact or discrete version. Depending on $\kappa$, or on $a$ defined in \eqref{defa}, this involves 
discrete amenable groups $G_a$, in particular the well-known Baumslag--Solitar ones. The associated 
$C^*$-algebras $\U_a$ can be viewed as a deformation of the two-torus. They are different when $a$ varies 
within the rational numbers (of zero Lebesgue measure) 
because of the structure of the underlying dynamical system. 
For transcendental values of $a$, which are dense in $\R_+$ and of full Lebesgue measure, all these algebras 
are isomorphic to each other.

Due to the exponential growth of $G_a$, we have proved 
that the algebras $\U_a$ are not only quasidiagonal but also residually finite dimensional. 
They admit different spectral triples: the ones which are quasi-equivalent to the left regular representation 
and are never $p$-summable but only $\theta$-summable, \textsl{i.e.} they are of ``infinite metric dimension". 
This situation reminds us of the passage from non-relativistic to relativistic quantum mechanics: in quantum field 
theory, the $\theta$-summability (and not the $p$-summability) naturally occurs due to the behaviour of 
$\Tr(e^{-tH})$ (when $t\rightarrow 0$) where $H$ is the Hamiltonian (or $\DD^2$), see for instance \cite{CHKL}.

The other faithful representations can generate fancy spectral triples which can have arbitrary summability (or 
``dimension") depending on the algebraic properties of the real parameter $a$, 
but are in fact degenerate to some extent. It is also not entirely clear what the topological content of these 
unbounded Fredholm modules is (i.e. whether they correspond to nontrivial elements of $K$-homology). 
The dimension of these spectral triples is unrelated to the number of coordinates defining the 
$\kappa$-deformed Minkowski spaces.

The nonexistence theorem, though powerful, does not preclude the possible existence of a genuine, 
non-degenerate, nontrivial spectral geometry on the $\kappa$-deformation spaces presented here, they only 
restrict the possible algebra representations that could be used in the construction.

This shows how delicate the notion of spectral or metric dimensions of  $\kappa$-Minkowski space is, and how 
subtle its analysis through noncommutative geometry.

\section*{Acknowledgments}

We thank Alain Connes, Michael Puschnigg, Adam Skalski and Shinji Yamashita for helpful discussions or 
correspondence. B. I. and T. S. acknowledge the warm hospitality of the Institute of Physics at the Jagiellonian 
University in Krakow where this work was started under the Transfer of Knowledge Program ``Geometry in 
Mathematical Physics".

\end{document}